\documentclass[times, 10pt]{article} 
\usepackage{graphicx,times,amsmath, amssymb}
\usepackage[T1]{fontenc} 
\usepackage{fixltx2e} 
\usepackage{caption}
\usepackage{float}
\usepackage{subcaption}
\usepackage{flushend}
\providecommand{\keywords}[1]{\textbf{\textit{Keywords---}} #1}

\begin{document}
\title{Structural Vulnerability Analysis of Electric Power Distribution Grids}

\author{Yakup Ko\c{c}$^{1,}$\thanks{Corresponding Author,~~Email: \texttt{Y.Koc@tudelft.nl},  Address: Jaffalaan 5, 2628BX Delft, the Netherlands, Phone: +31 (0)15 27 88380}~~Abhishek Raman$^2$~~Martijn Warnier$^1$~~Tarun Kumar$^2$\\
 $^1$Systems Engineering Section,\\
 Faculty of Technology, Policy and Management,\\
 Delft University of Technology, the Netherlands\\
 $^2$Smarter Energy Group,\\
 Department of Industry and Solutions,\\
 IBM T. J. Watson Research Center, Yorktown Heights, New York, USA\\
 }
\date{}

\maketitle
\thispagestyle{empty}

\begin{abstract}

Power grid outages cause huge economical and societal costs. Disruptions in the power distribution grid are responsible for a significant fraction of electric power unavailability to customers. The impact of extreme weather conditions, continuously increasing demand, and the over-ageing of assets in the grid, deteriorates the safety of electric power delivery in the near future. It is this dependence on electric power that necessitates further research in the power distribution grid security assessment. Thus measures to analyse the robustness characteristics and to identify vulnerabilities as they exist in the grid are of utmost importance. This research investigates exactly those concepts- the vulnerability and robustness of power distribution grids from a topological point of view, and proposes a metric to quantify them with respect to assets in a distribution grid. Real-world data is used to demonstrate the applicability of the proposed metric as a tool to assess the criticality of assets in a distribution grid.

\end{abstract}


\section{Introduction}
\label{sec_Introduction}


Among other critical infrastructures, the electric power grid plays a crucial role for the daily life in modern societies. This is not only because of the importance of electric power in the daily life, but also because of the dependency of other critical infrastructures on electric power~\cite{Eeten2011}. The continuous availability of electric power is of key importance for daily chores. Careful and diligent operations at the grid level ensure the transmission and distribution of electrical power with the highest possible reliability. Yet, electric power delivery disruptions do occur, causing huge economical and societal cost~\cite{Chen2001, Hines2009}. The analysis of the U.S. Energy Information Administration (EIA) data on power outages (of 100 MW or more) between 1991 and 2005 reveals that the increase in the number of outages in the North America has been increasing exponentially~\cite{Amin2008}. Also, disruptions in the electric distribution grid are responsible for 80\% to 95\% of customer electricity unavailability~\cite{Provoost2009}.

Power distribution grids are complex systems delivering electric power to millions of customers. In a typical distribution area, electric power is distributed as a result of the interaction of thousands of components, some of which are electrical assets (e.g. transformers, cables, switches), and others being physical assets (e.g. poles, insulators) which help support this propagation. In contrast to transmission grids that have a mesh structure, distribution grid networks have a radial structure, implying a minimum level of redundancy in the network structure. This minimum redundancy makes distribution networks more vulnerable to disturbances and external forces, calling for a complete power system security assessment of the distribution grids from various perspectives.

The vulnerability analysis of power grids can be classified as the conventional vulnerability analysis and the structural vulnerability analysis~\cite{Bompard2012}. The conventional vulnerability analysis requires complete operational and topological data as well as the engineering models in power systems. Yet, the boosting complexity of the determination of operational and topological states due to the increasing size of large-scale power grids challenges the conventional vulnerability analysis~\cite{Salmeron2014} of these systems. On the other hand, the strong relationship between the topology and physical behaviour of power systems makes the structural vulnerability analysis a promising alternative.  The structural vulnerability analysis is a complementary tool to the conventional vulnerability analysis rather than substitutionary. Moreover, the structural vulnerability analysis is also useful to understand the global properties of power grids affecting their local behaviours~\cite{Bompard2012}.

 The difference in the structure of power transmission and distribution grids affects the process in which failure propagates across the different layers of the grid. On the transmission side, the loss of one single component does not result in topological disconnection. However, it might trigger a cascade of successive failures in the form of line overloads based on capacity constraints, voltage and frequency level instabilities, and hidden failures in the protection devices, resulting in the eventual disconnection and impairment of the grid. Consequently, analysing the vulnerability of a transmission grid also requires the incorporation of the impact of power flow into account, thus suggesting that considering a purely topological approach would result in an incomplete analysis~\cite{Koc2013,Koc2013_2}. On the other hand, in the case of the distribution grid, because of its typical radial-like structure, the loss of one single component might potentially result in topological disconnection of a certain geographical region of the distribution grid. Due to the strong dependence of the distribution grid robustness to the underlying topology, the assessment of the distribution grid robustness from a \emph{topological} point of view is a promising approach to gain additional insight into the system intrinsics and behaviour.
 
Assessing power system vulnerabilities from a topological point of view requires a system level approach to capture the topological interdependencies in the system. Recent advances in the field of network science~\cite{Barabasi1999, Buldyrev2010, Huang2012, Watts1998} reveal the promising potential of complex networks theory to investigate power grids vulnerability at a system level. Accordingly, this paper models an electrical power distribution grid as a (directed) graph in which the nodes represent the electrical and physical assets in the system, while edges model the logical information about connections between these assets. 


Power system \emph{security} assessment investigates the ability of a system to provide service under unexpected operating conditions (e.g. contingencies). The power system \emph{vulnerability} indicates the sensitivity to threats (i.e. malicious attacks) and disturbances (e.g. random failures) that possibly limit the ability of the system to provide the intended services~\cite{Koc2014}. As opposed to vulnerability, \emph{robustness} refers to the ability of a system to perform the intended task under unforeseen disturbances. This paper focusses on the \emph{robustness} of an asset with respect to supply availability, and relates it to the ability of an asset to be connected to sources (for supply availability) from a \emph{topological} point of view.

The rest of this paper is organized as follows: Section~\ref{sec_RelatedWork} gives an overview of existing work on power system vulnerability assessment, and positions this work. Section~\ref{Section_RedVsRob} qualitatively discusses the difference between the traditional system safety concept redundancy and robustness. Section~\ref{Section_Metric} introduces the proposed metric Upstream Robustness. Section~\ref{Section_NumericalAnalysis} applies the proposed metric on real-world use cases to demonstrate its applicability for asset criticality assessment, while Section~\ref{Section_Conclusion} provides a conclusion and a discussion on future work.

\section{Structural Vulnerability Analysis of Power Distribution Grids}
\label{sec_RelatedWork}
Electrical power distribution security is an active field of research. The importance of the electrical power in daily life attracts many researchers to analyse the safety of electrical power delivery from various angles. Most of the existing efforts assesses grid safety from a reliability engineering perspective~\cite{Chertkov2011_2, Chertkov2011, Grainger1994, Gert2014, Yang2007}: researchers perform a quantitative analysis to estimate the system reliability performance based on component reliability values. The result of the analysis are the reliability indices indicating the ability of the system to deliver power to the load points. The reliability indices can be defined for individual load points, or for the overall system. Different approaches can be used to compute the reliability of a system including Reliability block diagrams, Markov methods, Petri nets and Monte Carlo Simulations~\cite{Rausand2004}. 

One important subject in power system vulnerability analysis is to identify the critical components in a power system~\cite{Dehghanian2012, Hilber2008, Hilber2007, Schneider2006, Schwan2007, Setreus2012}. The reliability indices (as a result of a reliability analysis) do provide reliability performances of the individual components, or overall system, however, they do not quantify the contribution of each component to the system reliability (i.e. criticality). To determine the criticality of a component, e.g. a sensitivity analysis is performed to relate the system reliability performance to the reliability performance of individual components~\cite{Birnbaum1968, Hilber2008}. In such an analysis, a component is critical for a system if a small change in the reliability of the component results in a comparatively large change in system reliability~\cite{Setreus2012}.

This approach from the \emph{reliability} perspective provides a detailed and useful analysis on system reliability, accounting for important aspects such as age, condition, and individual failure probabilities of the assets. However, it does not provide any explicit insight in the topological vulnerabilities of the components in the system. Moreover, it encounters various challenges including complexity of the computational methods and collecting accurate reliability data such as information on the material, age, and failure history of the asset which are not always available to the grid security analysts~\cite{Setreus2011}.

One complementary way to assess power system safety is structural robustness (or vulnerability) analysis, that is mainly performed from a Complex Networks Theory perspective. In such an analysis, a system is modelled as a graph, and metrics and concepts from Complex Networks Theory are deployed to statistically analyse topological characteristics of these modelled systems~\cite{Negeri2015}. In this way, the topology of a system is related to the (operational) performance of the system, so that the operation and design of these systems can be adjusted for a higher performance of the system.    

Researchers assess networked systems from a topological perspective in various fields including data communication networks~\cite{Sterbenz2013}, water management systems~\cite{Yazdani2012}, and transportation systems~\cite{Derrible2010}, and electrical power systems~\cite{Sole2008, Rosato2007, Bompard2010, Koc2014_2, Koc2014, Casals2007, Hines2009}. Some of these studies on power grids~\cite{Pagani2011, Sanchez2012, Wang2010} statistically investigate the topological properties of a power grid (such as degree distribution of clustering coefficient~\cite{Mieghem2011}) to relate its topology to the existing network models (e.g. small-world~\cite{Watts1998}, or scale-free~\cite{Barabasi1999}). A significant part of these studies on Complex Network Analysis of power grids~\cite{Sole2008, Rosato2007, Bompard2013, Koc2014_2, Bilis2013} investigate the relationship between the topology and the performance of the system. Relying on these analysis, to quantify and exploit this relationship between topology and performance, various metrics are designed/proposed. These metrics can be used for various purposes including vulnerability analysis of components (or of overall system), network design purposes, and critical component identification. Some of these studies propose \emph{extended} topological metrics that reflect the electrical properties of the power grid~\cite{Bompard2010, Hines2009, Koc2014_2, Koc2014}, while most of them~\cite{Crucitti2004, Albert2004, Rosato2007, Moreno2003, Sole2008} characterize the power grids in terms of classical topological metrics, such as Betweenness Centrality~\cite{Mieghem2011}.


Whereas most of the existing work on analysing power systems from a topology point of view focus on high-voltage power \emph{transmission} grids, recent studies analyse the structure of medium-, and low-voltage power \emph{distribution} systems~\cite{Negeri2015, Pagani2011}. Negeri et al.~\cite{Negeri2015} investigate the impact of topology of a distribution grid on its operational performance, while Pagani et al.~\cite{Pagani2011} asses the influence of the topological structure of a distribution grid on the cost of decentralized power trading. Despite these studies that statistically assess the power distribution grid topology, to the best of our knowledge, there has been no attempt on designing measures to quantitatively assess the robustness of an electrical \emph{distribution} grids from a topological point of view.

 The contribution of a component to the overall system security is determined by (i) the reliability performance (affected by say the type of the materials, ambient conditions, and age of the asset) and (ii) the connectivity (position relating to the topology) of the asset. Accordingly, a complete security assessment of an electric power distribution grid requires analysing both of the factors separately to assist the grid security analysts to identify the vulnerabilities and the asset managers to determine the criticality of assets in the system. As complementary to the probabilistic approach from the reliability point of view, this paper focuses on the assessment of the power system security from a \emph{topological} point of view, and proposes a metric, \emph{Upstream Robustness}, to quantify the robustness of an asset with respect to supply availability in a electric power distribution grid.

\section{Redundancy vs. Robusness}
\label{Section_RedVsRob}


The robustness of power \emph{distribution} grids heavily depends on the topology of the grid, mainly because of the minimum level of the redundancy in the system. Due to this lack of topological redundancy, unlike in the power transmission grid, the loss of a single component can potentially result in topological disconnection of a sizable part of the grid from the sources and thus the disruption of the power delivery to the corresponding part of the grid.

Therefore, measuring and managing redundancy in power distribution networks is of key importance. The redundancy of a component in a distribution grid is \emph{traditionally} measured in terms of the number of alternative paths to that component from the sources. This number of alternative paths represents the number of different ways to reach an asset (and ultimately customers). The presence of multiple alternative paths (characterized by redundancy) implies a more robust asset since even in case of a failure of one path due to loss of a single component along the path, the asset is still supplied power through other alternative paths, preventing it from being single sourced and vulnerable to single point failures in the system. 

However, the existence of multiple alternative paths to an asset does not necessarily protect it from being affected by single point failures. This is especially the case when such alternative paths have common components between them. Even though multiple paths exist between the asset and source(s), the failure of a common component still disconnects the asset from the sources. Hence, as much as the \emph{redundancy} (i.e. the number of paths), the \emph{quality} of this redundancy with respect to the exact configuration and involved components, also has a crucial role in determining the robustness of the asset with respect to supply availability. The number of alternative paths to an asset evaluates the redundancy of the asset to receive electric power and the level of disjointness and the length of these paths determine the quality of this redundancy. 


\begin{figure}\centering
\includegraphics[width=0.45\textwidth]{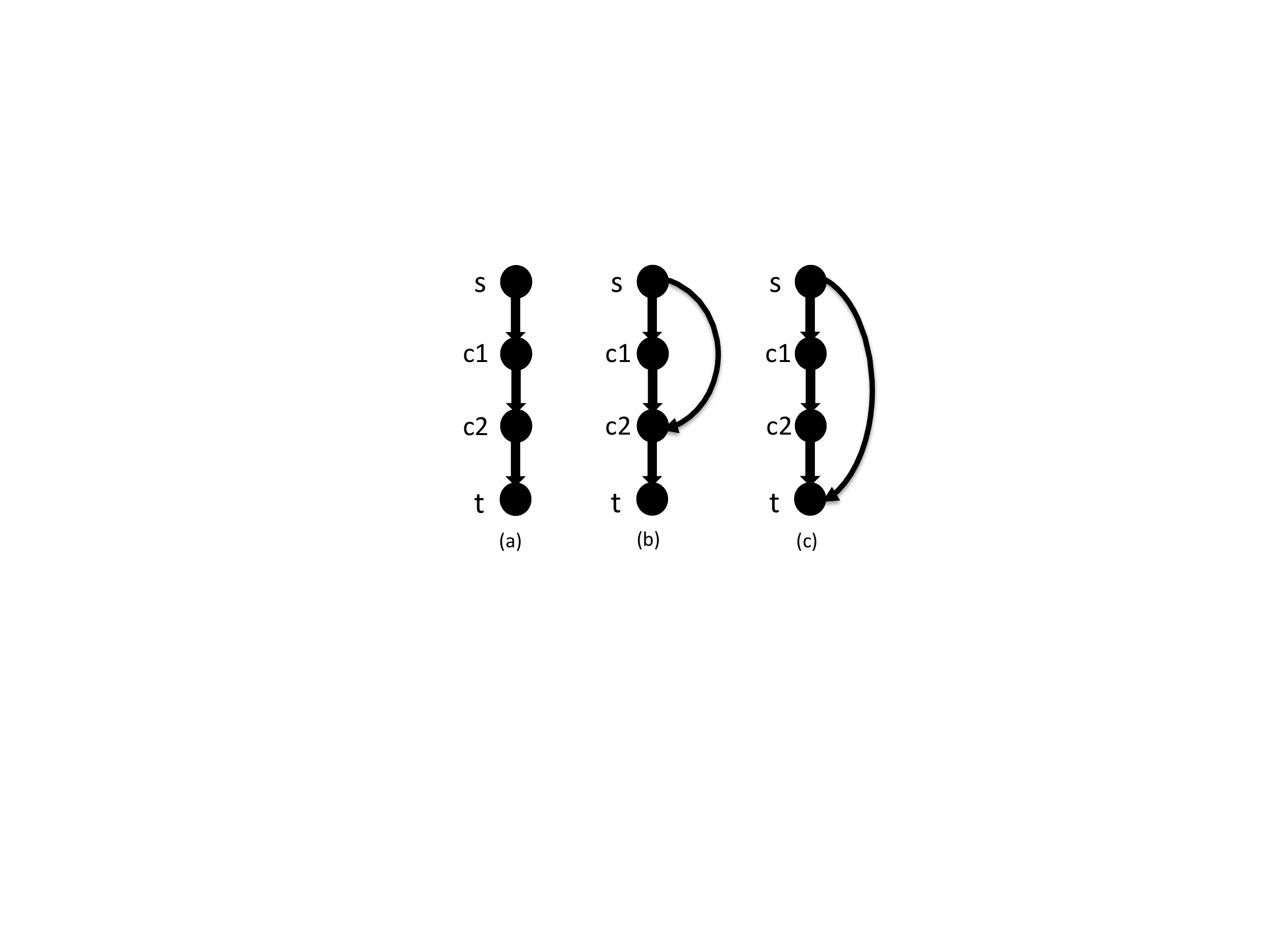}
\caption{Robustness versus redundancy}
\label{fig:RobustnessVSRedundancy}
\end{figure}

 Fig.~\ref{fig:RobustnessVSRedundancy} illustrates the relationship between the robustness and the redundancy along with its quality. In the first topology in  Fig.~\ref{fig:RobustnessVSRedundancy}.a, only one path exists between the target node $t$ and the source $s$. The target node is not redundant, and any failure of the components in the upstream results in disconnection of node $t$ from the source. In the second and third topology, two alternative paths exist between the target node $t$ and the source node $s$. These paths are partially disjoint in the second topology, while they are completely disjoint in the last topology. As a result of two completely disjoint paths, in the second topology, all of the components are backed up and no single component failure causes disconnection of $t$ from the source $s$, as opposed to the second topology in which the single failure of node $c2$ results in the disconnection of $t$ from $s$ across both available paths.    

In addition to the disjointness of alternative paths, the length of these paths (in terms of number of the assets in these paths) is another aspect contributing to the quality of the redundancy in a power distribution grid. Each additional component in a path introduces additional uncertainty and vulnerability by increasing the number of the components on which $t$ depends to be connected to the sources in the system. 

To quantify the robustness of an asset, the \emph{Upstream Robustness} captures the redundancy of the network along with its quality by accounting for (i) the number of alternative paths (from sources) to the asset, and (ii) the disjointness and (iii) the length of these alternative paths.  



 \section{Metric: Upstream Robustness}
\label{Section_Metric}

The proposed metric Upstream Robustness $R_{ups}$ relies on two main concepts; the inter-path independency and intra-path independency. This section introduces these concepts, explains the computational methodology, and elaborates on how these concepts are combined to quantify the upstream robustness of a given asset.

\subsection{Inter-path independency}
\label{SubSection_PI_inter}

The Upstream Robustness of an asset $t$ is a weighted sum of the inter-path independency values of all possible paths to $t$. The \emph{inter-path independency} $PI_{t,i}^{inter}$ of a path $i$ quantifies the independency of the path $i$ with respect to the other alternative paths to $t$.

In this work, the independency of a path with respect to the neighbouring paths corresponds to the disjointness between these paths. Given a set of alternative paths $P_{t}=\{P_{t,1},P_{t,2},...,P_{t,m}\}$ to $t$ that is obtained by e.g. a Breadth-First Search algorithm~\cite{Korf1993}, quantifying the disjointness of these paths requires capturing the impact of (i) the number of common components between these alternative paths, and (ii) the length of these paths to determine the fraction of the common components within a path.

\begin{figure}\centering
\includegraphics[width=0.5\textwidth]{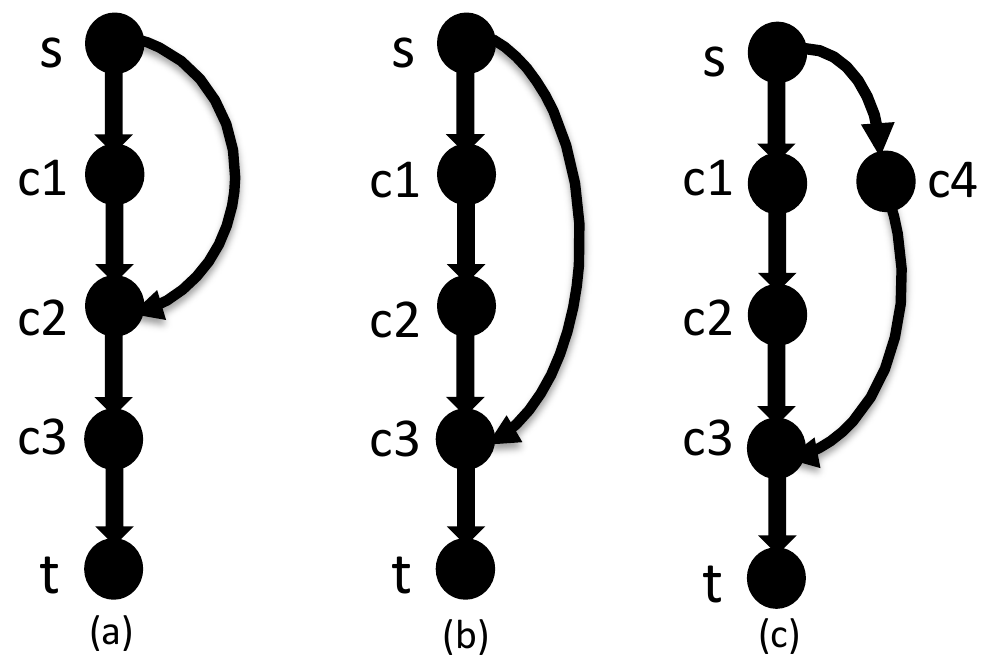}
\caption{Inter-path Independency conceptual explanation}
\label{fig:interPathIndependency}
\end{figure}

 Fig.~\ref{fig:interPathIndependency} shows three conceptual topologies for an intiutive discussion on the impact of the number of common components, and the length of a path on the disjointness. In all topologies, a target asset $t$ is connected to a single source $s$ through different configurations. In all the topologies, two different alternative paths exist between $s$ and $t$; e.g. for Fig.~\ref{fig:interPathIndependency}.a, $P_{t,1}=\{c1,c2,c3\}$ and  $P_{t,2}=\{c2,c3\}$.

 In  Fig.~\ref{fig:interPathIndependency}.b, $P_{t,1}$ and $P_{t,2}$ have higher disjointness levels compared to $P_{t,1}$ and $P_{t,2}$ in Fig.~\ref{fig:interPathIndependency}.a, because in Fig.~\ref{fig:interPathIndependency}.b, $P_{t,1}$ and $P_{t,2}$ have fewer components in common (1 component compared of 2). Additionally, in Fig.~\ref{fig:interPathIndependency}.c, the length of $P_{t,2}$ increases while the number of common components with $P_{t,1}$ remains the same; hence the fraction of the common components within $P_{t,2}$ decreases. As a result of this decrease, the disjointness of $P_{t,2}$ in Fig.~\ref{fig:interPathIndependency}.c increases compared to the disjointness of $P_{t,2}$ in Fig.~\ref{fig:interPathIndependency}.b.




For a set of alternative paths $P_{t}$ between the source(s) and a target node $t$, the computational methodology for the inter-path independancy $PI_{t,i}^{inter}$ of a path $i$ to an asset $t$ requires determining the \emph{Universe} of the components between the source(s) and the asset $t$, evaluating the \emph{frequency} of the occurrence of each asset in these alternative paths, and relying on this frequency value, computing the \emph{score} of each component in the universe. The normalized summation of scores of assets in a path results in the inter-path independancy of the path.




The collection of the components in the alternative paths between the source(s) and $t$ comprise the universe $U_{t}$ of the components. $U_{t}$ is the union of all alternative paths in $P_{t}$.


\begin{equation*}
U_{t}=\left\{P_{t,1} \cup P_{t,2} \cup ,...,\cup, P_{t,m}\right\}
\end{equation*}

The frequency $f_{c}$ of a component $c$ in $U_{t}$ is determined by evaluating the number of occurrence of $c$ in all alternative paths in $P_{t}$. The total number of occurrence of $c$ in these paths gives the frequency of occurrence $f_{c}$ of $c$. 


\begin{equation*}
f_{c} = \begin{cases}
f_{c}+1 &\text{if $c\in P_{t,i}$}\\
f_{c} &\text{if $c\not\in P_{t,i}$}
\end{cases}
\end{equation*}

The score $s_{c}$ of $c$ defines how much the component belongs to a path and it is computed as the inverse of the frequency $f_{c}$: 

\begin{equation}
s_{c}=\frac{1}{f_{c}}
\end{equation} 
 
 For a completely disjoint path, all of its components belong to the path completely; each component has a score of one. The score of a component decreases as it is shared between multiple alternative paths. 

$PI_{t,i}^{inter}$ of a path $P_{t,i}$ is determined by summing up the score of each component in $P_{t,i}$, and normalizing it with the summation of the maximum possible scores of the components in $P_{t,i}$. The summation of the component scores is maximized for the case all the components have frequency and score of 1, i.e. when a path is completely disjoint. In that case, the summation of the scores equals the number of the components in $P_{t,i}$. 

\begin{equation}
PI_{t,i}^{inter}=\frac{1}{L_{t,i}}{\sum_{i \in P_{t,i}} s_{i}} 
\end{equation} 

where $L_{t,i}$ is the the number of the assets in the path $P_{t,i}$, and computed as the cardinality of $P_{t,i}$:

\begin{equation}
L_{t,i}=|P_{t,i}|
\label{Eq:pathCardinality}
\end{equation} 

$PI_{t,i}^{inter}$ quantifies how disjoint the paths are to each other in $P_{t}$. The maximum value of $PI^{inter}$ of a path is one, for a completely disjoint path. $PI^{inter}$ of a path decreases as more and more components are shared with other neighbouring paths.

\subsection{Intra-path independency}
\label{SubSection_PI_intra}

The \emph{intra-path independency} $PI_{t,i}^{intra}$ of a path $P_{t,i}$ to an asset $t$ is an asset-specific metric relating to the number of assets in $P_{t,i}$ that have to be functioning for $t$ receiving supply from the dedicated source $s$. Each additional asset in a path between $t$ and s increases the dependency of $t$ on other assets to reach $s$, accordingly decreasing the robustness of $t$ with respect to the supply availability. Hence, the number of the components in a path to a given asset is inversely proportional to the robustness of the given asset to be connected to the source $s$. The intra-path independency conceptualizes this effect of additional components in a path on the robustness with respect to supply availability. Quantifying the impact of the number of the components in a path $P_{t,i}$ requires first determining the length $L_{t,i}$ (see Eq.~\ref{Eq:pathCardinality}) of the path in terms of number of the components in the path excluding the source and target nodes. Intra-path independency $PI_{t,i}^{intra}$ of the path $P_{t,i}$ is a function of the length of the path $P_{t,i}$.
 
\begin{equation}
\label{Eq:intraDep}
PI_{t,i}^{intra}=\frac{1}{L_{t,i}+1}
\end{equation} 
 
\noindent In the denominator in Eq.~\ref{Eq:intraDep}, the $+ 1$ expression accounts for the effect of the failure of the target node itself. In other words, for a target node $t$ to function as expected, all of the components in its path from the source $s$ along with itself need to not fail and work as expected.

\subsection{Upstream robustness}
\label{Subsection_upstream robustness}
After computing the inter-path and intra-path independencies, two values are assigned to each path. The product of these values for a path gives the individual contribution of the path to the overall upstream robustness of the asset. Finally, the Upstream Robustness of an asset $t$ can be computed by aggregating the individual contributions of each path to the overall robustness of $t$.

\begin{equation}
R_{ups}^t=\sum_{i=1}^{m}PI_{t,i}^{inter}PI_{t,i}^{intra}   
\end{equation} 

\noindent where $m$ stands for the number of paths from the source(s) to $t$.

\section{Use Case: Asset Criticality Assessment}
\label{Section_NumericalAnalysis}


Together with the de-regulation of the electricity markets, distribution grid operators face the challenging task of attaining the delicate balance between providing adequate power reliably and its economical ramifications. The expectations from the customers and government bodies on high supply availability enforces utility companies to deploy effective methods to reduce their cost while maintaining good service levels when it comes to delivering this power reliably across the network. Addressing this dilemma requires assessing the criticality of components and prioritizing them for effective decision-making on maintenance strategies and on investment plans. This section demonstrates the ability of the Upstream Robustness $R_{ups}$ metric to determine the criticality of assets in a power distribution grid and thus to aid with making effective decisions. 

In a criticality analysis relying on $R_{ups}$, the criticality of an asset $c$ relates to the extent in which $c$ contributes to the upstream robustness of the assets that are downstream of it (i.e. the assets that are connected to a source via asset $c$). Accordingly, the criticality of an asset $c$ is computed as the drop in the collective robustness of the other assets in the system upon removal of $c$. A sensitivity analysis is performed to compute the criticality $C_{down}^{c}$ of a component $c$ in a distribution grid/network $G$:

\begin{equation}
\label{Eq:SensitivityAnalysis}
C_{down}^{c}=\frac{R_{ups}^{G}-R_{ups}^{\grave{G}}}{R_{ups}^{G}}
\end{equation} 

\noindent where $\grave{G}$ is the weakened system that is obtained by removing the asset $c$ from the original system $G$. The $R_{ups}^{G}$ is the Upstream Robustness of the original network $G$:

\begin{equation}
\label{Eq:NetworkRobustness}
 R_{ups}^{G}=\frac{1}{N} \sum_{i=1}^{N} R_{ups}^{c_{i}}
\end{equation} 

\noindent where $N$ corresponds to the number of the components which are load points of the system, e.g. transformers that are potentially connected to customers, or batteries that are used to store electric power in micro grids. 

\begin{figure*}\centering
\includegraphics[width=1.1\textwidth]{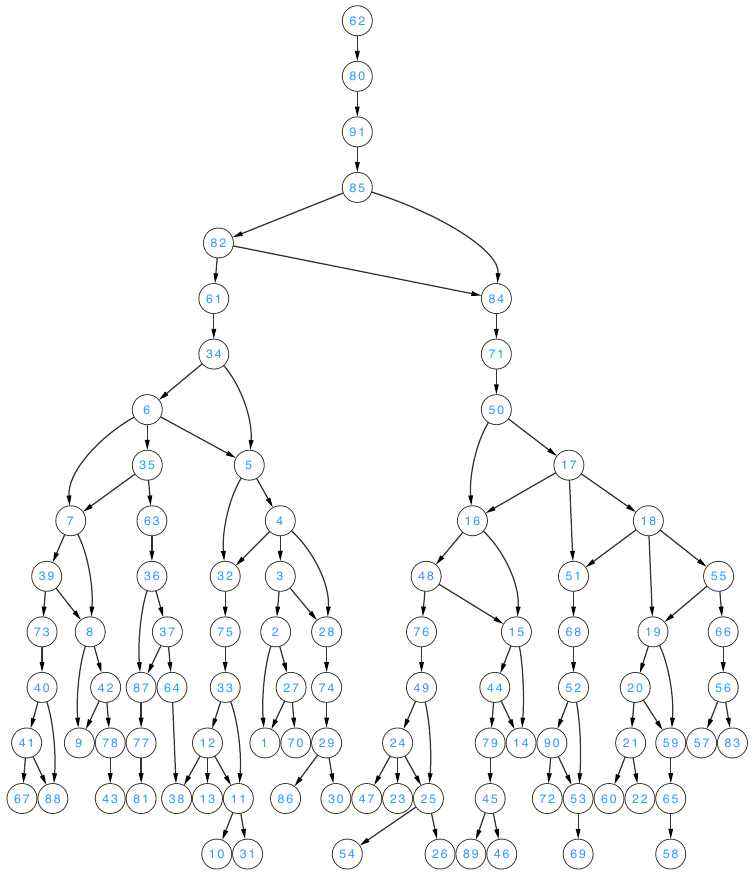}
\caption{Graph representation of a part of the real world power distribution grid substation.}
\label{fig:ToyNetwork}
\end{figure*}

\begin{figure*}\centering
\includegraphics[width=1\textwidth]{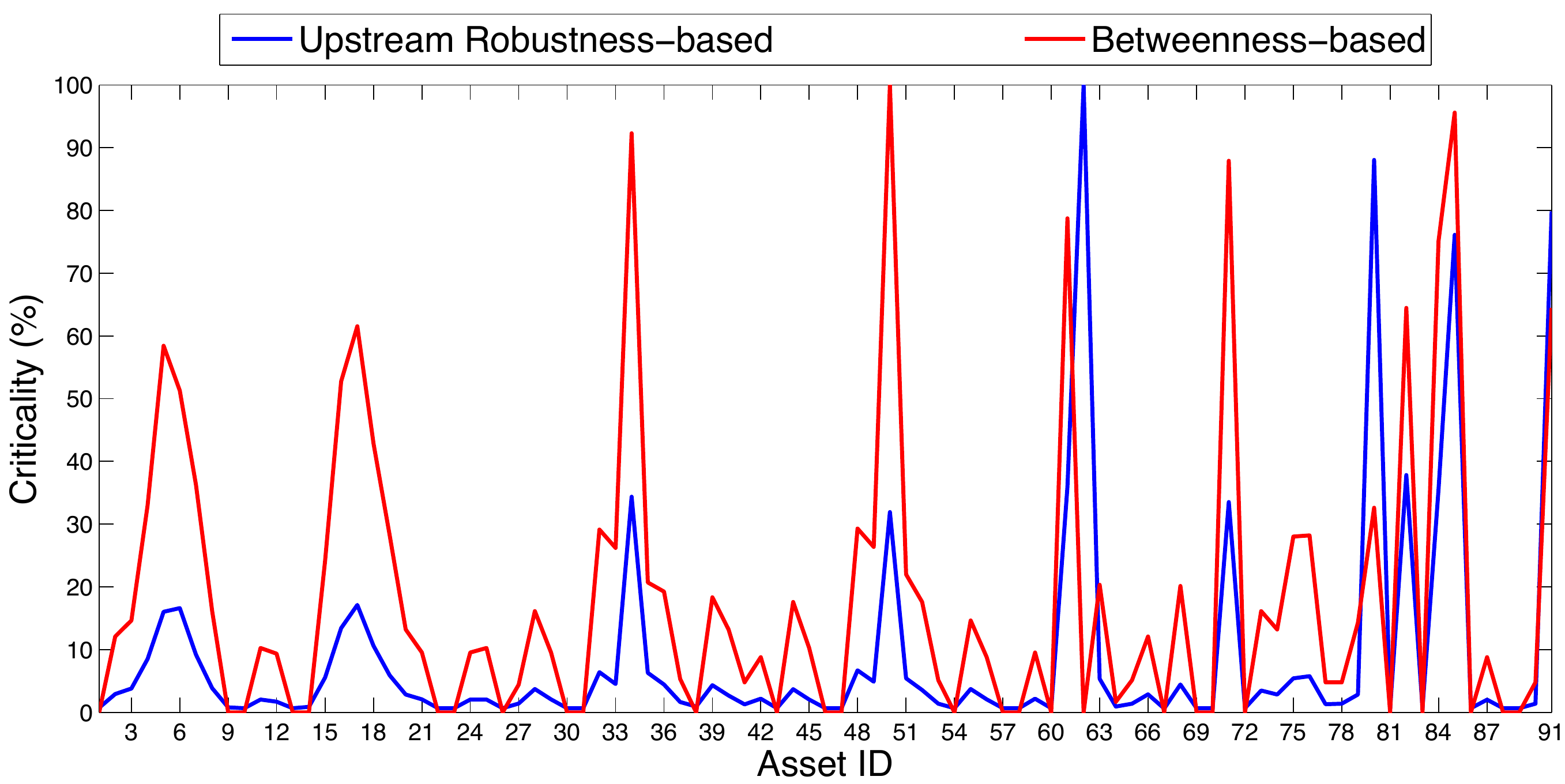}
\caption{Asset criticalities of the sub-network of the real world power distribution grid substation. Blue line gives the result of criticality analysis by $R_{ups}$ based on sensitivity analysis (See Eq.~\ref{Eq:SensitivityAnalysis}), while the red line corresponds to the normalized Betweenness Centrality values of assets. Normalized Betweenness values are obtained by normalizing all the Betweenness values of assets with the largest Betweenness value of the assets (i.e. Betweenness value of asset with ID 50). This is done to map all the Betweenness values in [0,1].}
\label{fig:ToyNetworkCriticalities}
\end{figure*}

To demonstrate the applicability of $R_{ups}$ as a measure to assess the criticality of the components, a sub-grid of an actual electrical power utility company's distribution grid is considered. Fig.~\ref{fig:ToyNetwork} illustrates a hierarchical graph representation of the sub-grid. In this graph, nodes correspond to the assets such as transformers and cables, while the edges model the logical connections between these assets. The network consists of 91 assets comprised of electrical and physical assets, and supplied by one single source.

The criticality of assets in Fig.~\ref{fig:ToyNetwork} are determined by a sensitivity analysis (See Eq.~\ref{Eq:SensitivityAnalysis}) relying on $R_{ups}$. At the same time, for comparison purposes, the criticality of assets are also determined based on \emph{Betweenness Centrality}. The Betweenness Centrality of an asset $c$ $B_{c}$ relates to the number of shortest paths in the system that traverse the asset $c$. 

\begin{equation}\label{Betweenness}
B_{c}=\sum_{\substack{s,t \in N }}{\frac{\sigma_{st}(c)}{\sigma_{st}}}
\end{equation}

\noindent where $\sigma_{st}(c)$ is the number of shortest paths passing through asset \textit{c}, while $\sigma_{st}$ is the total number of shortest paths in the grid topology. A relatively larger $B_{c}$ corresponds a larger number of shortest paths through the asset $c$, implying a higher criticality of $c$ in the system. Therefore, the Betweenness Centrality is traditionally used to quantify the criticality of components in complex networks, also applied on power grids~\cite{Albert2004, Crucitti2004, Moreno2003}. Fig.~\ref{fig:ToyNetworkCriticalities} shows asset criticality values based on Upstream Robustness $R_{ups}$ and Betweenness Centrality $B$ (blue line for criticality based on $R_{ups}$ and red line for $B$), while Tab.~\ref{tab:TopCriticalAssetsToyNetwork} gives the 15 most critical assets identified by these metrics.


\begin{table}
\caption{The top 15 critical assets identified by Upstream Robustness $R_{ups}$ and Betweenness Centrality $B$.}
\centering
\begin{tabular}{ l c l c}    
 \hline                  
  Asset ID &  $R_{ups}$- based criticality (\%) &  Asset ID &  $B$- based criticality (\%)\\
 \hline  \hline    
62 &	100.00 & 50 &	100.00\\
80 &	88.05 & 85 &	95.60\\
91 &	79.87 & 34 &	92.31\\
85 &	76.11 & 71 &	87.91\\
82 &	37.82 & 61 &	78.75\\
61 &	35.95 & 84 &	75.09\\
84 &	35.50 & 82 &	64.47\\
34 &	34.36 & 91 &	64.46\\
71 &	33.51 & 17 &	61.54\\
50 &	31.92 & 5 &	58.42\\
17 &	17.10 & 16 &	52.75\\
6	 &16.60 & 6 &	51.28\\
5   &	16.03 & 18 &	42.86\\
16 &	13.42 & 7 &	36.26\\
18 &	10.59 & 4 &	32.97\\
  \hline  
\end{tabular}
\label{tab:TopCriticalAssetsToyNetwork}
\end{table}


Fig.~\ref{fig:ToyNetworkCriticalities} illustrates the similarity between the asset criticality values by the $R_{ups}$-based and $B$-based approach. In Fig.~\ref{fig:ToyNetworkCriticalities}, most of the times, both metrics identify the same assets as critical (e.g. assets with ID 34, 50, and 71), and non-critical (e.g. 54, 67, and 81). Also Tab.~\ref{tab:TopCriticalAssetsToyNetwork} shows the top 15 assets according to $R_{ups}$ and $B$ are mostly in line: 13 of the 15 assets are common in these lists (although with different rankings). However, in both Fig.~\ref{fig:ToyNetworkCriticalities} and Tab.~\ref{tab:TopCriticalAssetsToyNetwork}, one big difference is obvious in criticality identification of nodes with ID 62 and 80. The asset criticality analysis based on $R_{ups}$ show that the asset with ID of 62 and 80 are the most critical assets for the robustness of the system, while they are considered not critical by $B$. These results collectively show that the two metrics capture similar properties of a topology to some extent, however there are differences between them too.

The Betweenness Centrality is a topological measure that is widely used to assess network characteristics in complex networks. Although a topological approach is appropriate to assess the power \emph{distribution} grid (See the discussion in Sec.~\ref{sec_Introduction}), purely topological generic metrics such as Betweenness Centrality fail to capture certain topological characteristics of power grids, mainly for two reasons~\cite{Bompard2009}. First, the Betweenness Centrality does not make any distinction in the type of buses in the system. However, in power grids, each bus can be categorized depending on its function as generation, transmission and load buses. The electric power is transmitted from the generation buses to load points through intermediate (i.e. transmission) components. The goal of a power grid is delivering electric power from the generation to the load points. Therefore, \emph{only} the paths between the generation and the load points matter, rather than shortest paths between \emph{any} pair of nodes. Second, $B$ considers only the shortest path between a pair of nodes, e.g. a source and a load point. However, accounting for all paths between these buses along with their quality (i.e. disjointness and length of these paths) is of key importance to measure the robustness in the system (See discussion in Sec.~\ref{Section_RedVsRob}). The Upstream Robustness quantifies the redundancy in the system along with its quality. These two main differences between $R_{ups}$ and $B$ results in different assessment of component criticality analysis in the system.

In Fig.~\ref{fig:ToyNetwork}, two sub-areas are visible, and these areas are fed by one single source (the asset with ID 62). Accordingly, in Fig.~\ref{fig:ToyNetworkCriticalities} and in Tab.~\ref{tab:TopCriticalAssetsToyNetwork}, $R_{ups}$ identifies the asset with ID 62 as the most critical component in the system with a criticality value of 100\%. This is because this component is the only source supplying the network, and loss of this component results in the loss of all other components in the system. However, $B$ quantifies the criticality of node with 62 as 0, since it does not distinguish between the bus types, failing to capture the importance of a source in a power grid. 

In Fig.~\ref{fig:ToyNetworkCriticalities}, according to the analysis based on $R_{ups}$, 27 assets have a criticality of lower than 1 \%. These components are "leaf" components, i.e. no downstream components are attached to them (e.g. the component with the ID of 10 in Fig.~\ref{fig:ToyNetwork}). Accordingly, the failure of these components does not negatively affect the supply availability of any other components except for themselves. This is captured by the criticality analysis based on $R_{ups}$. Also $B$ identifies these components as non-critical, since these assets are topologically not central, i.e. no shortest paths traverse these assets.

In the component criticality assessment based on $R_{ups}$ (See Tab.~\ref{tab:TopCriticalAssetsToyNetwork}), the first 3 components (after the source node) are the assets that are in very close vicinity of the source (assets with ID 80, 91, and 85) and their loss results in de-energizing of the rest of the system. Consequently, $R_{ups}$ spot them as the most critical components after the source component in the system. After that, gradually, as the components move farther away from the source, the criticality of assets also drops, since the number of the components that depend on them reduces as well. Visually, in Fig.~\ref{fig:ToyNetwork}, the loss of any of the assets with ID 82, 61, or 34 disconnects all the assets in the left hand side of Fig.~\ref{fig:ToyNetwork}, while the failure of any of the  assets with ID 84, 71 or 50 disengages all the assets in the right hand side of the Fig.~\ref{fig:ToyNetwork}. The criticality of these assets is captured by the criticality analysis based on $R_{ups}$, and these assets are identified as the top critical assets in the system, verifying the effectiveness of criticality analysis intuitively.

\begin{figure*} [htb]
\centering
\includegraphics[width=1\textwidth]{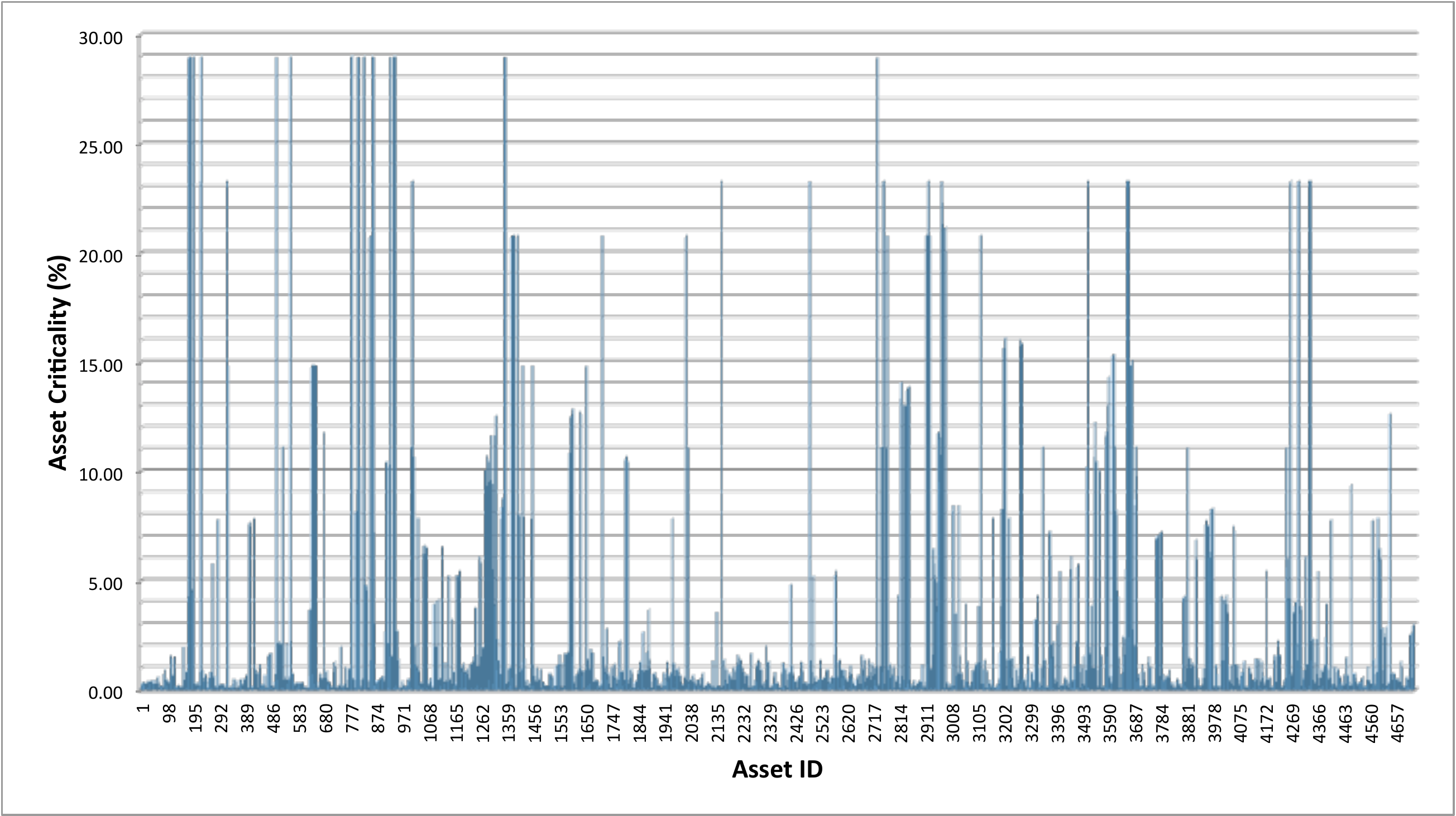}
\caption{The criticality of the assets in the \emph{Substation A}.}
\label{fig:LilyNetworkCriticalities}
\end{figure*}

To assess whether the proposed metric $R_{ups}$ is globally applicable and scalable, a criticality analysis is a performed for a real-world substation region of the same electrical power utility's territory. Because of the sensitivity of such information, we anonymized the asset IDs and will refer to this substation region as \emph{Substation A}. 

\emph{Substation A} consists of 4713 assets that are fed by 6 different feeders (i.e. sources) in the system. 1314 of these assets are transformers, while the rest are assets such as cables, breakers, fuses, switches, reclosers and support structures (e.g. poles). The entire Substation A is modelled as a graph, the network robustness value is determined based on Eq.~\ref{Eq:NetworkRobustness}, and the criticality of each asset is assessed based on Eq.~\ref{Eq:SensitivityAnalysis}. Fig.~\ref{fig:LilyNetworkCriticalities} shows the results.  

Fig.~\ref{fig:LilyNetworkCriticalities} suggests the existence of a small subset of components in the system with relatively high criticality, and a significant fraction of the components with much lower (than 1\%) importance to the system. These critical components are geographically spread over the substation area, implying that not only the distance to the sources, but also the topological positions of these components matter in determining the criticality of the assets in the system.  

Fig.~\ref{fig:LilyNetworkCriticalities} shows that, as opposed to the criticality analysis in Fig.~\ref{fig:ToyNetworkCriticalities}, sources in the Substation A do not have 100\% criticality for the system. The region covered by Substation A is fed by multiple sources. Each source feeds a part of the network, and some parts of the network are fed by multiple sources. Consequently, a loss of a single source affects only that part of the network which isn't redundant; the remainder of the network can still be supported by the 5 other sources in the system. 

\begin{figure}\centering
\includegraphics[width=0.7\textwidth]{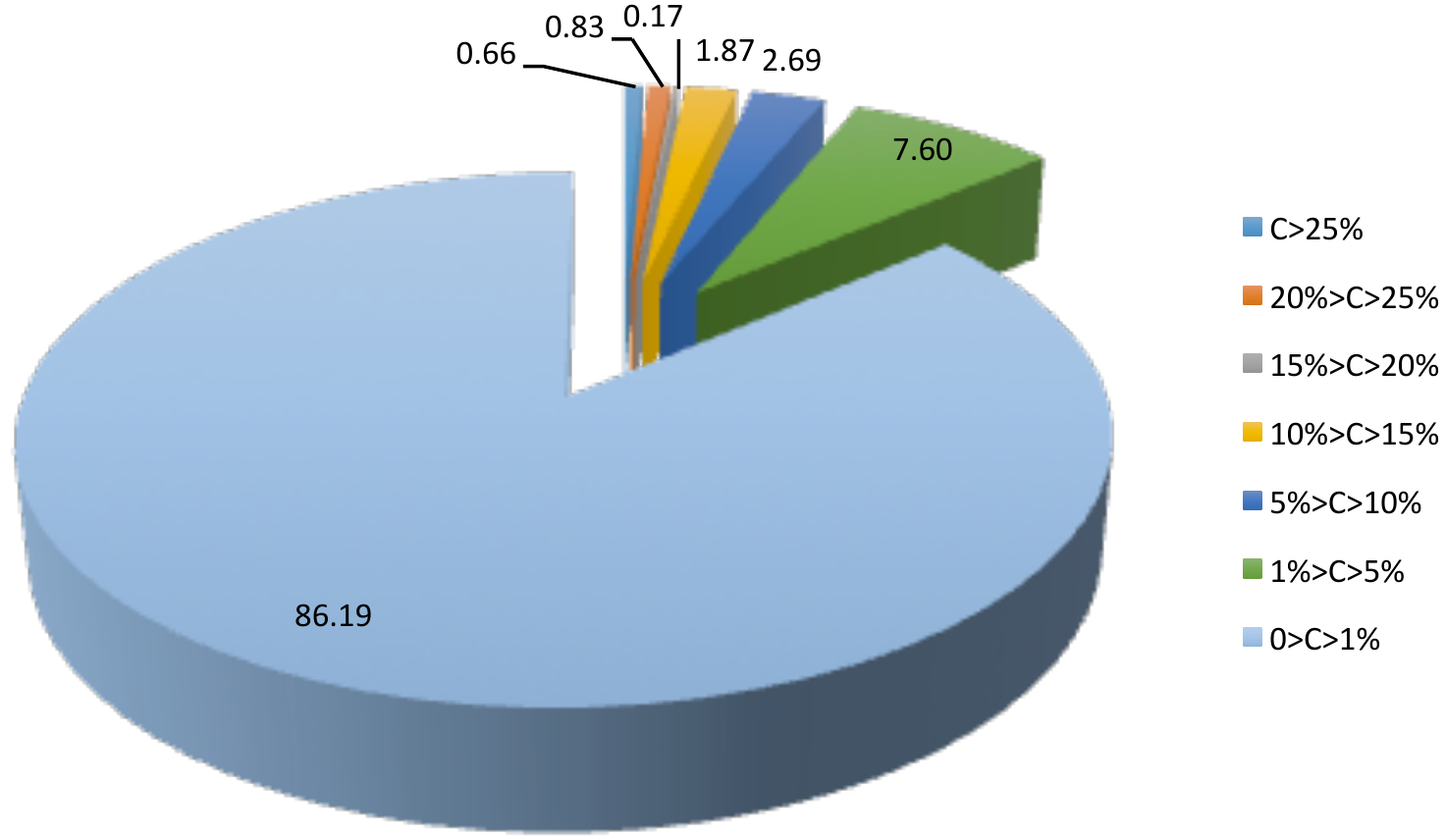}
\caption{Asset criticality distribution in the \emph{Substation A}. 86.19\% of the assets in the system have a criticality less than 1\%, while only 0.66\% of the assets have a criticality of larger than 25\%}
\label{fig:CriticalityDistributionPieChart}
\end{figure}

Fig.~\ref{fig:CriticalityDistributionPieChart} illustrates how the criticality of the assets in \emph{Substation A} (See Fig.~\ref{fig:LilyNetworkCriticalities}) is distributed over all assets. In \emph{Substation A}, only 31 out of nearly 5000 components have a criticality larger than 25\%, 39 more have a criticality that is larger than 20 \%, and in total 166 components with a criticality more than 10\%. Hence, less than 2\% of all components have a criticality larger than 20\%, and less than 4\% of the components in the system have a criticality more than 10\%. On the other hand, more than 80\% of the assets in the system have a criticality of less than 1\% suggesting that removal of these components have a very minor impact on the robustness of other assets in the system. 

As in every complex system, including the power distribution grid, the relative importance of the components in the system is highly skewed and non-uniform. A very small subset of the components is significantly critical for the system, while a greater portion of components have relatively small criticality values. In Fig.~\ref{fig:LilyNetworkCriticalities}, the criticality analysis based on the Upstream Robustness identifies these critical components in \emph{Substation A} effectively, showing the value of such metrics in assisting asset managers for appropriate decision-making on investment plans.


\section{Conclusion and Discussion}
\label{Section_Conclusion}

Utility companies typically assess the robustness of their network at different levels of grid abstraction. The first method corresponds to a circuit level analysis where traditional reliability engineering techniques and concepts are focussed on. A circuit is traced from its source and feeder in the substation towards downstream components. While doing so, the condition of the involved assets are estimated and used as a proxy for individual asset robustness and overall system reliability. A more topological driven approach is employed when it comes to criticality assessment. The whole network is divided into protective zone regions based on the presence of protective devices such as fuses, reclosers and sectionalizers. The count of customers is aggregated in each of these protective zones, and used as a KPI (Key Performance Indicator) to denote the effective deployment of protective devices in the grid. Both these techniques fail to in capture the presence of alternate paths of electrical distribution in the grid which is enabled by devices such as switches and open points. This combined with the knowledge of path length and path disjointness results in a much more useful and accurate way of computing network robustness, which has been the focus of our work.

This paper proposes the Upstream Robustness $R_{ups}$ to quantify the robustness of an asset against disturbances to receive supply from the sources. The computation of the upstream robustness of an asset $c$, $R_{ups}^c$, requires measuring the redundancy along with the quality of the redundancy of the topology between the asset $c$ and the source(s) in the network. The Upstream Robustness achieves this by accounting for three main aspects of the topology: (i) the number of alternative paths (from sources) to the asset, (ii) the level of disjointness of these alternative paths, (iii) the number of the assets in these alternative paths. 

The alternative paths to an asset is determined by a \emph{Breadth First Search} algorithm. The disjointness of these alternative paths to each other is modelled by \emph{inter-path independency} while the impact of the length of these paths on the assets robustness is captured by \emph{intra-path independency}. Combining these (inter)dependencies, the Upstream Robustness of the asset is computed.


The proposed metric is used to assess the criticality of the assets in a given distribution grid. $R_{ups}$ is applied on the real-world data of a distribution substation network to investigate the criticality of the assets in the system. These results are compared with the results from a traditional complex networks metric, Betweenness Centrality, and the differences between two approach are discussed. Experimental results confirm the effectiveness of the proposed metric $R_{ups}$ as a measure for asset criticality analysis to assist asset managers in appropriate decision-making regarding the investment and maintenance planning of the grid. Yet, it should not be the only approach to consider when making investment decisions to improve a grid. As is true with most power distribution grid utilities, factors like customer priority (such as assets serving important industrial customers or critical services like hospitals), regulatory requirements, capacity constraints, health of individual assets, and economic benefits should also be considered and given their due weightage. That said, because of the large number of assets involved to make the system work, not only does this technique provide a new lens to assess assets, but also a way to rank them when the other factors are equal.

This paper focuses on the robustness assessment of traditional power distribution grids. Future work focusses on making the proposed metric applicable for the Smart grid case. Accounting for the smart power distribution grids requires slight adjustments to the proposed metric. For instance, the introduction of prosumer concept in smart grids enables bidirectional flow rather than a unidirectional power flow as it is now in a traditional power distribution grid. Incorporating the impact of bidirectional power flow into the proposed approach requires determining all possible paths between an asset and sources with the assumption of undirected graph modelling of the grid rather than directed graph. The future work will focus on incorporating all such aspects so that the metric can also be deployed to assess the robustness of smart grids. Additionally, the proposed metric will also be applied on various relevant problems including evaluation of the right locations for adding assets (such as cables, transformers, and batteries) for future network expansion planning.


 \subsection*{{\bf Acknowledgements}}

{
 This work is partially funded by the NWO project \emph{RobuSmart: Increasing
   the Robustness of Smart Grids through distributed energy
   generation: a complex network approach}, grant number
 647.000.001.}


\vspace*{\fill}

\end{document}